\documentclass[aps,prb,twocolumn,amsmath,amssymb,superscriptaddress,floatfix,showpacs]{revtex4-1}
\usepackage{times,amsmath,amsfonts,amssymb,mathrsfs,graphics,graphicx,color,comment,bm}
\usepackage{natbib}
\usepackage{epsfig}
\usepackage[next]{inputenc}
\usepackage[pdfstartview=FitH]{hyperref}
\usepackage{graphicx}

\newcommand{\dg}{{\dagger}}

\newcommand{\nn}{\nonumber}          
\newcommand{\la}{\langle}          
\newcommand{\ra}{\rangle}          
\newcommand{\pbar}{\bar{p}}

\newcommand{\bk}{{{\bf{k}}}}

\newcommand{\kb}{\textbf{k}}

\begin{document}

\title{Phase diagram of $J_1-J_2$ transverse field Ising model on the checkerboard lattice: a plaquette-operator approach}

\author{M. Sadrzadeh}
\affiliation{Department of Physics, Sharif University of Technology, P.O.Box
11155-9161, Tehran, Iran}
\email{marzieh_sadrzadeh@physics.sharif.edu}
\homepage{http://spin.cscm.ir/}
\author{A. Langari}
\affiliation{Department of Physics, Sharif University of Technology, P.O.Box
11155-9161, Tehran, Iran}
\affiliation{Center of excellence in Complex Systems and Condensed Matter
(CSCM), Sharif
University of Technology, Tehran 1458889694, Iran}
\affiliation{Max-Planck-Institut f\"ur Physik komplexer Systeme, 01187 Dresden,
Germany}
%\email{langari@sharif.edu}
%\homepage{http://sharif.edu/~langari/}

\date{\today}

\begin{abstract}
We study the effect of quantum fluctuations by means of a transverse magnetic field ($\Gamma$) on the
antiferromagnetic $J_1-J_2$ 
Ising model on the checkerboard lattice, the two dimensional version of the pyrochlore lattice. 
The zero-temperature phase diagram of the model has been obtained by employing a plaquette operator approach (POA).
The plaquette operator formalism bosonizes the model, in which a single boson is associated
to each eigenstate of a plaquette and the inter-plaquette interactions define an effective Hamiltonian.
The excitations of a plaquette would represent an-harmonic fluctuations of the model, 
which lead not only to lower the excitation energy compared with a single-spin flip but also 
to lift the extensive degeneracy in favor of a plaquette ordered solid (RPS) state,
which breaks lattice translational symmetry, in addition to a unique collinear phase for $J_2>J_1$. 
The bosonic excitation gap vanishes at the critical points to the N\'{e}el ($J_2 < J_1$) and collinear ($J_2 > J_1$) ordered phases, which defines the critical phase boundaries. At the homogeneous coupling ($J_2=J_1$) and its close neighborhood, the (canted) RPS state, established from an-harmonic fluctuations, lasts for low fields, $\Gamma/J_1\lesssim 0.3$, which is followed by a transition to the quantum paramagnet (polarized) phase at high fields. The transition from RPS state to the N\'{e}el phase is either a deconfined quantum phase transition or a first order one, 
however a continuous transition occurs between RPS and collinear phases.
\end{abstract}
\pacs{75.10.Jm, 75.30.Kz, 64.70.Tg}

\maketitle

\section{Introduction\label{introduction}}
Frustrated magnetic systems imply large degenerate classical configurations as a groundstate subspace,
which could lead to novel phases and exotic features like emergent magnetic
monopoles in spin ice \cite{Castelnovo:2008}.
Quantum fluctuations as perturbations may select one of these degenerate states as a unique quantum ground-state 
of the system representing unusual ordering.
Besides the magnetic properties, which are described by models of frustrated systems,
such models mimic some features of unconventional superconductivity in terms of resonating
valence bond (RVB) phase \cite{Anderson:1973} on the triangular lattice  
\cite{Moessner:2000,Moessner:2001PRL,Moessner:2001PRB,Misguich:2008}
, governed by quantum dimer model (QDM) \cite{Rokhsar:1988}. The RVB scenario has received a great
impact to elucidate the plaquette RVB phase in the s=1/2 honeycomb $J_1-J_2$ Heisenberg model
\cite{Albuquerque:2011,Mosadeq:2011}, which is justified by the two-dimensional approach of 
density matrix renormalization group, recently \cite{Zhu:2013,Ganesh:2013PRL}.
It gives the impression that the plaquette type ordered phase is a result of strong correlation
and frustration, which has also been observed in the square lattice 
\cite{Zhitomirsky:1996,Isaev:2009,Wenzel:2012,doretto2014plaquette,gong2014plaquette}. According to Ref.~\cite{doretto2014plaquette}, which shows that a plaquette phase is stable
in the range of parameters, where a spin-liquid phase has been reported \cite{gong2014plaquette,jiang:2012}, it is interesting to investigae the spin-liquid phase within a plaquette operator approach.

\begin{figure}
\centerline{\includegraphics[width=85mm]{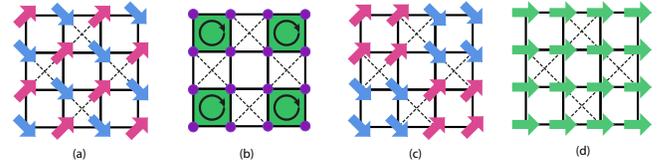}}
\caption{(color online) Schematic representations for various phases of the $J_1-J_2$ 
transverse field Ising model on the checkerboard lattice. The solid and dashed lines 
are $J_1$ and $J_2$ bonds, respectively. (a) The (canted) N\'{e}el state,( b) the (canted) RPS state, 
(c) the (canted) collinear state and (d) the quantum paramagnet (polarized) state.}
\label{checkerboard}
\end{figure}

The three-dimensional pyrochlore lattice is a fascinating example of geometrically frustrated lattices,
which has a two-dimensional (2D) version called checkerboard lattice (see Fig.~\ref{checkerboard}). 
Quantum Heisenberg model has been widely studied on checkerboard lattice,
where degeneracy of the groundstate is lifted toward a unique non-magnetic ordered state, which
is called a plaquette ordered phase or a plaquette valence bond solid (pVBS).
This state does not break SU(2) symmetry and is gapped but breaks the space symmetry 
of the lattice
\cite{Palmer:2001,Sindzingre:2002,Brenig:2002,Canals:2002,Fouet:2003,Tchernyshyov:2003,Bernier:2004,Starykh:2005,chan:2011,Bishop:2012}.

However, the reduction of symmetry from SU(2) to Z2 renders
the antiferromagnetic Ising model on the checkerboard lattice as a prototype of frustrated systems,
which gives interesting features.
At the isotropic coupling $J_2=J_1$, the Ising model on the checkerboard lattice
has an extensive degenerate 
ground state defined by {\it ice-rule} manifold, i.e. 'two-in-two-out' on crossed squares, which imitates  a classical spin liquid known as square ice. 
Quantum fluctuations lift the degeneracy of the manifold toward a single magnetic \cite{Moessner:2001PRB}
or non-magnetic plaquette ordered state \cite{Shannon:2004,Moessner:2004}.
In Ref.~\cite{Shannon:2004}, the ice-rule manifold is mapped to the spin configurations of quantum six-vertex model.
The quantum fluctuations of a weak in-plane XY-term are considered in terms of the second order perturnation
on the ice-rule manifold, which leads to cyclic cluster terms that can be modeled by a QDM of flippable plaquettes,
which stabilizes
a plaquette phase at zero chemical potential \cite{Shannon:2004,Syljuasen:2006}.
However, in this article we show explicitly the existence of a resonating plaquette solid state in terms of
its corresponding order parameter, which will be introduced. Moreover, we indicate the region, 
where an RPS state is being formed in the neighborhood of $J_2=J_1$ of our phase diagram.

We study a general transverse field Ising model (TFIM) on the checkerboard lattice:
\begin{equation}
\label{eq1}
	\mathcal{H}=		J_1\displaystyle\sum_{\langle i,j \rangle}{S_i^zS_j^z}
							+J_2\displaystyle\sum_{\langle\langle i,j \rangle\rangle}{S_i^zS_j^z}
							-\Gamma\displaystyle\sum_{i}{S_i^x}~
\end{equation}
where $J_1>0$ is the nearest neighbor coupling, $J_2>0$ is the diagonal coupling on crossed squares, $\Gamma$ is the 
strength of transverse magnetic field and $S^{x,z}$ refer to {\it x} and {\it z} components of spin-1/2 operators 
on the vertices of the lattice. In the absence of transverse field $\Gamma$, as well as exponentially degenerate groundstate (with the system size) of the isotropic case $J_2=J_1$, there is exponentially degenerate groundstate with the linear size of the system for $J_1<J_2$ (the collinear phase) 
and a unique groundstate (although with a $Z_2$ degeneracy) for $J_1>J_2$ (the N\'{e}el phase).

Recently, the transverse field Ising model on the $J_1-J_2$ checkerboard lattice
has been studied within linear spin-wave theory (LSWT)\cite{Henry:2012}.
The phase diagram 
consists of three phases, N\'{e}el ordered for low magnetic field and $J_2<J_1$, highly degenerate collinear
phase at low field and $J_2>J_1$ and fully polarized phase for high transverse fields ($\Gamma$).
Based on harmonic fluctuations considered in Ref.~\cite{Henry:2012}, the boundary 
between N\'{e}el and collinear phases is at $J_2=J_1$ for $0 \leq \Gamma \lesssim 0.7$
without an indication of an RPS phase, which is a witness for the break down of LSWT.
Moreover, the border for the polarized-N\'{e}el and polarized-collinear phase transitions can not
be determined accurately within LSWT due to strong quantum fluctuations, leading to instabilities close to
the phase boundaries \cite{Henry:2012}. 

The first clue to solve the problem is to employ the proper building blocks, which 
incorporate the correct ingredients of the ground state structure and the elementary excitations of the model.
At zero field and in the intermediate regime i.e. for $2/3 < J_2/J_1 <4/3$, 
where the role of frustration is important,
a plaquette flip excitation
has lower energy than a single spin-flip \cite{Henry:2012}, which suggests that the true 
excitations of the model is governed by a plaquette flip that is a representation of an-harmonic fluctuations
(of the original spin model). Moreover, the zeroth-order calculations of the ground state energy
immediately justify that a single plaquette background gives lower value than a single particle classical background.
We implement a plaquette-operator approach (POA)\cite{Marcelo:2007,Ganesh:2013}, which is an
extension of the bond-operator theory \cite{Sachdev:1990} to obtain the
zero temperature phase diagram of $J_1-J_2$ TFIM on the checkerboard lattice, accurately.
We explicitly find the quantum phase boundary for paramagnet-N\'{e}el and 
paramagnet-collinear transitions, where the excitation energy of bosonic quasi-particles vanishes as 
the onset of a Bose-Einstein condensation. The corresponding phase diagram is presented in
Fig.~\ref{phasediagram}. Moreover, we show that an-harmonic
fluctuations lift the extensive degeneracy of the collinear phase to form a unique quantum 
state of collinear order.
In addition, the phase transition between N\'{e}el and collinear phases only appears at zero field ($\Gamma=0$)
and $J_2=J_1$ while for small field region a (canted) RPS phase fills the phase diagram. 
The RPS phase breaks the translational symmetry of the lattice, which has twofold degeneracy. The increment of 
transverse field causes a transition from the (canted) RPS phase to the quantum-paramagnet one.

Our paper is organized as follows: In sec.~\ref{poa} we describe the plaquette operator approach applied for TFIM
on the checkerboard lattice. We obtain and discuss the POA results and compare them with LSWT 
ones in Sec.~\ref{results}. Finally, we summarize and conclude in sec.~\ref{conclusion}. 
Some details of our calculations of 
the groundstate energy, correlations, and the order parameters can be found in \ref{ap-a}, \ref{ap-b} and \ref{ap-c}, respectively.

\section{Plaquette Operator Approach \label{poa}}

The plaquette operator approach is an extension of the bond-operator formalism \cite{Sachdev:1990},
where the bond is replaced by a cluster of spins, namely: plaquette.
In the bond-operator approach, a pair of spins -- a bond -- is treated exactly and a bosonic operator is associated
to each eigenstate of the bond. A condensation for the lowest (energy) boson is considered as the 
background configuration of the model. The effect of inter-bond interactions is taken into account perturbatively
in terms of boson operators, which defines the effective theory for the original spin model. 
The ground state energy is minimized self consistently, which includes the corrections caused by the quantum fluctuations
of quasi-particle bosons. To preserve the Hilbert space, a constraint has to be 
imposed on the boson occupation of each bond, i.e. the total occupation of all bosons on a single bond have to 
be equal to unity.

In the plaquette operator approach the system is divided to a set of individual plaquettes
that are shaded as one every two uncrossed squares of the checkerboard lattice
(see Fig.~\ref{checkerboard}-(b)).
The total Hamiltonian, Eq.~\ref{eq1}, is written as $\mathcal{H}=\mathcal{H}_0+\mathcal{H}_{int}$, where $\mathcal{H}_0$ denotes 
the Hamiltonian for the set of non-corner-sharing uncrossed squares, called plaquettes, and $\mathcal{H}_{int}$ 
represents the interaction 
between plaquettes. The plaquette Hamiltonian ($\mathcal{H}_0$) is solved exactly and its lowest eigenstate
is subjected to the Bose-condensation that defines the system background.
The elementary excitations are of a plaquette type, which are created as a consequence of
inter-plaquette interactions ($\mathcal{H}_{int}$) out of the Bose-condensated background.
The ground state energy is corrected by considering the inter-plaquette interactions,
which leads to the proper
quasi-particle excitations of the model that determine the critical phase boundaries at the
location of vanishing of the energy gap.

\subsection{A single plaquette}

\begin{figure}
\centerline{\includegraphics[width=45mm]{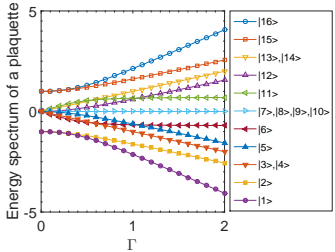}}
\caption{(color online) Energy levels, in units of $J_1$, of a single plaquette versus transverse field ($\Gamma$). The levels ($\vert3\rangle,\vert4\rangle$), ($\vert13\rangle,\vert14\rangle$) and ($\vert7\rangle,\vert8\rangle,\vert9\rangle,\vert10\rangle$) are degenerate. The bottom ($\bullet$) line is the unique groundstate of the plaquette.}
\label{spectrum}
\end{figure}

The Hamiltonian of a single plaquette is in the following form (the scale of energy is set by $J_1=1$):
\begin{equation}
{\mathcal{H}}_{P}=\displaystyle\sum_{\langle i,j \rangle}{S_i^zS_j^z}-\Gamma\displaystyle\sum_{i}{S_i^x}
\label{eq2}
\end{equation}
where $i,j=1,2,3,4$ are the indices of the four spins located on the corners of 
a plaquette (see Fig.~\ref{coupling}). Hence, 
$\mathcal{H}_0$ is a sum on all (non-corner sharing) plaquettes, i.e. $\mathcal{H}_0=\sum_P {\mathcal{H}}_{P}$.
The single-plaquette Hamiltonian is diagonalized exactly and its energy spectrum versus transverse field $\Gamma$
is plotted in Fig.~\ref{spectrum}. We see that the ground state of a plaquette is non degenerate except for $\Gamma=0$. 
Hence, in a non-zero transverse field and in the absence of interaction between plaquettes, 
all of the isolated plaquettes are in their unique groundstates. 

\begin{figure}
\centerline{\includegraphics[width=95mm]{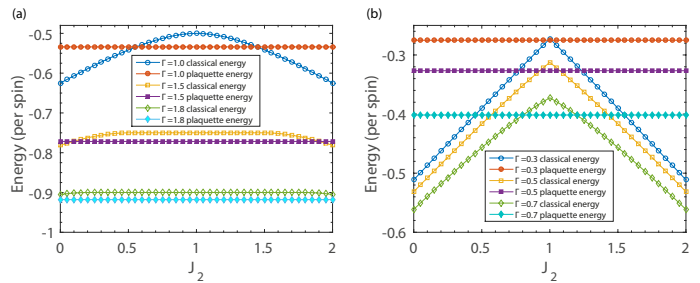}}
\caption{(color online) Ground state energy (per spin) versus $J_2$ at fixed values of the transverse 
magentic field. (a) $\Gamma=1.0, 1.5, 1.8$ and (b) $\Gamma=0.3, 0.5, 0.7$ .
For each value of $\Gamma$ the open symbol shows the classical ground state energy and the filled one corresponds
to the quantum plaquette order ($\mathcal{H}_{0}$).}
\label{comparebackground}
\end{figure}

The zeroth order approximation gives an impression on how a plaquette background ($\mathcal{H}_0$) would lead to 
a proper approximation for the ground state energy.
In Fig.~\ref{comparebackground} we have compared the ground state energy of the classical approximation
-- that has been used as the background configuration in LSWT \cite{Henry:2012}--
with the ground state energy of the quantum plaquette order ($\mathcal{H}_0$), which is employed
as a background in POA (the present work).
Accordingly, the following two facts can be deduced.
(i) The ground state energy of a plaquette is lower than the classical one for 
high-field values, Fig.~\ref{comparebackground}-(a). It shows that
POA is a high-field approach within the chosen plaquettes of Fig.~\ref{checkerboard}-(b). 
Thus, we expect to get reasonable 
excitations of the model for high field values and arrive at a gapless critical point
by reducing the transverse field for a fixed value of the exchange coupling $J_2$.
(ii) For the intermediate region of the exchange coupling, namely: $J_2 \sim J_1$, where
the frustration prohibits that all bonds being minimized classically, the dominant term is the
transverse field compared with the frustrated exchange ones even at low field values, see 
Fig.~\ref{comparebackground}-(b). It suggests that we should expect acceptable 
results for the low transverse fields close to the highly frustrated regime $J_1 \sim J_2$.

The inter-plaquette interactions excite the plaquettes to the higher eigenstates, which
reduces the probability of a single 
plaquette to be in its groundstate. The quantum fluctuations caused by the inter-plaquette interactions
modify the ground state energy and render the proper excitations of the model to govern the
critical boundaries.

\subsection{Interaction between plaquettes: a bosonic representation}

\begin{figure}
\centerline{\includegraphics[width=35mm]{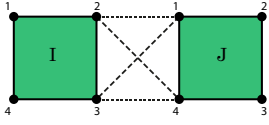}}
\caption{The interaction between two `nearest-neighbor' plaquettes, $I$ and $J$. The dotted and dashed lines are $J_1$ and $J_2$ couplings, respectively. 
}
\label{coupling}
\end{figure}

The inter-plaquette Hamiltonian, $\mathcal{H}_{int}=\mathcal{H}-\mathcal{H}_0$, which is
composed of Ising terms on the dotted and dashed links of Fig.~\ref{coupling}, does not commute with the plaquette 
Hamiltonian $\mathcal{H}_0$ that includes the transverse field. As a result, the inter-plaquette
interactions hybridize the ground state of a single plaquette with the corresponding excited eigenstates.
In order to take into account the effect of inter-plaquette interactions, we implement a bosonization formalism\cite{Ganesh:2013} 
similar to what has been introduced as bond-operator representation of 
spin systems \cite{Sachdev:1990,Isoda:1998,Langari:2006,Rezania:2008,Rezania:2009,Rezania:2010}. 
A boson is associated to each eigenstate $\vert u \rangle$ of a single-plaquette Hamiltonian 
such that the eigenstate is created by the corresponding boson creation operator $b_{I,u}^\dagger$ 
acting on the vacuum,
\begin{equation}
\vert u \rangle_I= b_{I,u}^\dagger \vert 0 \rangle, \hspace{5mm} u=1, \dots, 16,
\label{eq3}
\end{equation}
where $I$ denotes the plaquette label of a shaded square of Fig.~\ref{checkerboard}-(b). 
The bosonic operators $b_{u}^\dagger$ and  $b_{u}$ obey 
the known commutation relation $[b_{u}, b_{u}^\dagger]= 1$.

In the absence of inter-plaquette interactions, all of the isolated plaquettes are in their groundstates, 
$\vert 1 \rangle$. Therefore, a plaquette ordered state can be defined as a Bose-condensation of 
the groundstate bosons. 
We assign a Bose-condensation amplitude $\bar{p}_I$, 
\begin{equation}
 \bar{p}^2_I \equiv \langle b_{I,1}^\dagger b_{I,1} \rangle, 
 \label{eq4}
\end{equation}
which gives the probability of a single plaquette to be in its ground state. For simplicity and
within a mean-field level of approximation we consider 
\begin{equation}
b_{I,1} \equiv b_{I,1}^\dagger \equiv \bar{p},
\label{eq5}
\end{equation} 
for all plaquettes, which is equal to unity in the absence of inter-plaquette interactions.
However, taking into account the inter-plaquette interactions, 
the value of $\bar{p}^2$ reduces from its perfect plaquette ordering amplitude, i.e. $\bar{p}^2 \lesssim1$, 
giving rise to a non-zero occupation of other excited bosons, which defines an effective theory
for the interacting model (see Eq.~\ref{eq8}). To preserve the Hilbert space, we impose the 
constraint of unit boson occupation for each isolated plaquette, i.e.
\begin{equation}
 N\bar{p}^2 +\sum_{I,u=2}^{16} b_{I,u}^\dagger b_{I,u} =N,
\label{eq6}
\end{equation}
where N is the total number of shaded uncrossed squares in Fig.~\ref{checkerboard}-(b). 

The inter-plaquette interaction between two plaquettes which is shown by Ising terms on 
dotted and dashed lines of Fig.~\ref{coupling}, is called $H_{IJ}$ (details are given in \ref{ap-a}).
The state of two nearest neighbor plaquettes $I$ and $J$ in the absence of interactions, is given by $\vert I_u J_v \ra$,
indicating that the plaquette-$I$ is in the $\vert u \rangle$ state and plaquette-$J$ in $\vert v \rangle$.
The inter-plaquette interactions are considered in terms of the matrix elements of $H_{IJ}$ between two product states, i.e. $\la I_u J_v \vert H_{IJ} \vert I_s J_t \ra$.
%between two direct product states $\vert I_u J_v\ra$ and $\vert I_s J_t \ra$.
However, because of the Bose-condensation assumption of the ground state background, all other excited bosons will be present in
very dilute concentrations. Thus interactions between such dilute bosons are unlikely and
we only consider transitions between $u=1$ groundstate-bosons of each plaquette and the other excited 
bosons, i.e. we do not consider the matrix elements between excited bosons themselves. 
In other words, either $\vert u \ra$ or $\vert s \ra$ for 
plaquette-$I$ and either $\vert v \ra$ or $\vert t \ra$ for plaquette-$J$ are necessarily in the state $\vert 1 \ra$.
Hence, only terms proportional to $\bar{p}^2$  participate in the effective Hamiltonian, 
resulting in a quadratic bosonic form.

\begin{figure}
\centerline{\includegraphics[width=95mm]{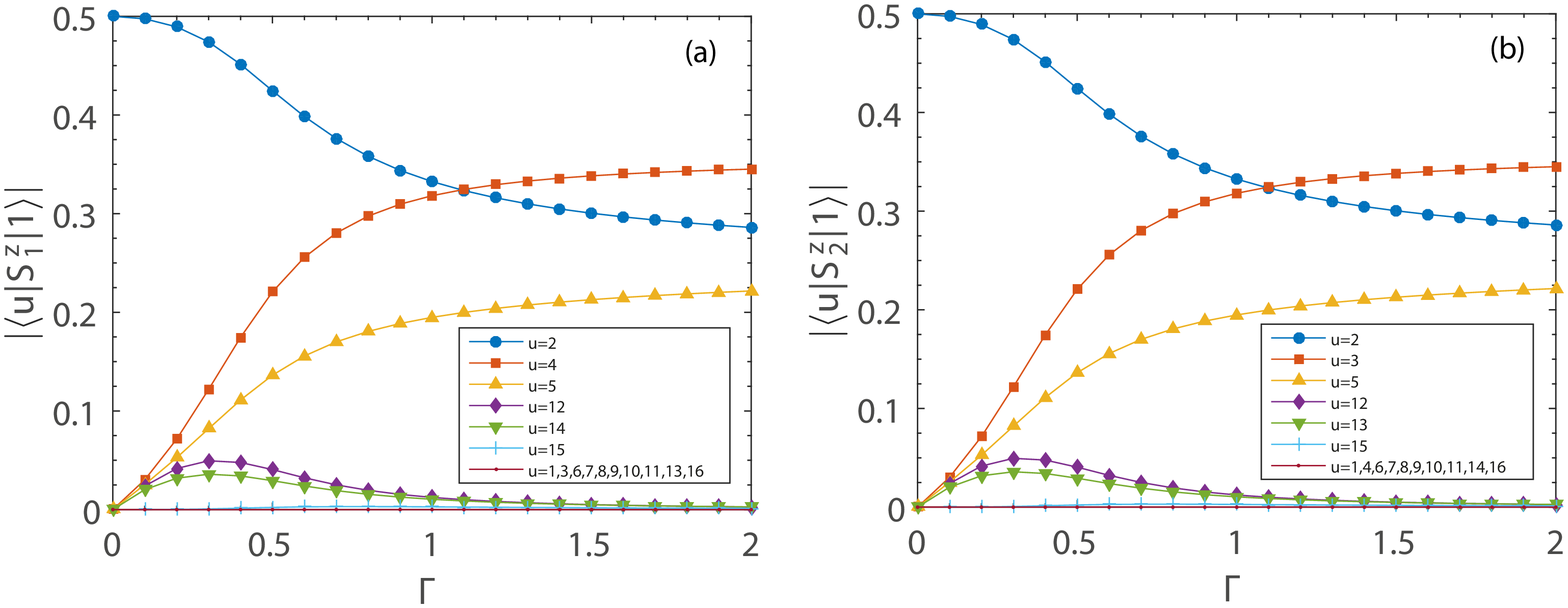}}
\caption{(color online) Transition amplitudes versus magnetic field between the ground state ($|1\rangle$)
and sixteen eigenstates ($|u\rangle, u=1, 2, 3, \dots, 16$) of a plaquette. 
(a)  $|\langle u | S^z_1 |1 \rangle|$ that is equal to $|\langle u | S^z_3 |1 \rangle|$.
(b) $|\langle u | S^z_2 |1 \rangle|$, which is equal to $|\langle u | S^z_4 |1 \rangle|$.}
\label{transition}
\end{figure}

The matrix element $\la I_u J_v \vert H_{IJ} \vert I_s J_t \ra$ is proportional to the 
transition amplitudes of two $S^z$ operators
\begin{equation}
 \la I_u J_v \vert H_{IJ} \vert I_s J_t \ra \sim 
 \sum_{\{i, j\}} \langle u | S^z_{i, I} |1 \rangle \times \langle v | S^z_{j, J} |1 \rangle,
 \label{eq7}
\end{equation}
where $\{i, j\}$ represents the interaction terms between two adjacent plaquettes (see Fig.~\ref{coupling}).
We have plotted the transition amplitudes in Fig.~\ref{transition}. It shows that transitions to only eight excited bosonic states are non zero with more
significant magnitudes for $u=2, 3, 4, 5$ (see also Table.~\ref{tab} in \ref{ap-a}).
Hence, we consider only the first four excited states 
of each plaquette that contribute in effective Hamiltonian. However, 
for low transverse fields $\Gamma\lesssim0.6$, the transition to the first excited state ($u=2$)
has a dominant role among the four mentioned excited states. 
Therefore, for low transverse fields in which we obtain an 
emergent RPS phase, we can reduce the number 'four' of excited bosons
to only 'one' boson (the first excited state of a plaquette). We use this simplified 
version of our approach in the appendices to obtain the groundstate energy, correlations and order parameters, analytically. 
However, the whole results of this paper are based on an effective Hamiltonian with four excited
bosonic states, described below.

\subsection{Effective Hamiltonian}

Finally, taking into account the inter-plaquette interactions,
the effective Hamiltonian of the system in the quadratic bosonic form, 
accompanied by the unit boson occupancy constraint via chemical potential $\mu$, reads as: 
\begin{eqnarray}
\nonumber\mathcal{H}&=&\sum_{I}\epsilon_{1}\bar{p}^2+ \sum_{I}\sum_{u}\epsilon_u b_{I,u}^\dagger b_{I,u}\\
\nonumber &&-\mu\Big[ N\bar{p}^2 +\sum_{I,u} b_{I,u}^\dagger b_{I,u} -N \Big]\\
\nonumber &&+\bar{p}^2 \sum_{\langle IJ\rangle}\sum_{u,v}\Big[\langle uv\vert H_{IJ}\vert11\rangle\:  
b_{I,u}^\dagger b_{J,v}^\dagger\\
&&+\langle u1\vert H_{IJ}\vert1v\rangle\:b_{I,u}^\dagger b_{J,v}+ H.c.\Big],
\label{eq8}
\end{eqnarray}
where $u$ and $v$ run over the four dominant excited bosonic states of the nearest neighbor plaquettes $I$ and $J$, 
respectively. 
It should be noticed that the Z$_2$ symmetry of the original Hamiltonian, Eq.~\ref{eq1}, is respected
in the effective Hamiltonian.
This is a consequence of the eigenstates of Eq.~\ref{eq2} that preserve the Z$_2$ symmetry and, thus,
all of the bosonic states participating in the effective Hamiltonian keep this symmetry.
The Hamiltonian is written in the momentum space and within a paraunitary Bogoliubov transformation \cite{Colpa:1978}
we arrive at the following diagonal form (see \ref{ap-a} for details),
%%%%%%%%%%%%%%%%%%%%%
\begin{eqnarray}
\nonumber \mathcal{H}&=& N\mu+N\bar{p}^2(\epsilon_1- \mu) -\frac{1}{2}N\sum_{u}(\epsilon_u-\mu)
\\  
&&+\sum_{\kb} \sum_{\nu=1}^{4}\big(\frac{1}{2} + \gamma^\dagger_{\kb,\nu}\gamma_{\kb,\nu}\big)\omega_{\kb,\nu}(\mu,\bar{p}),
\label{eq9}
\end{eqnarray}
%%%%%%%%%%%%%%%%%%%%%
where $\bk$ sums over the first Brillouin zone of a square lattice constructed from the centers of the 
shaded plaquettes of Fig.~\ref{checkerboard}-(b), $\omega_{\bk,\nu}$ defines the spectrum of quasi-particles of the interacting model and $\gamma^\dagger_{\bk,\nu}$ is the corresponding bosonic creation operator.
The constant term in Eq.~\ref{eq9} represents the ground state energy of the plaquette-ordered background, which is corrected due 
to the interaction between plaquettes and takes into account the zero point quantum fluctuations of plaquette type. The two parameters $\bar{p}$ and $\mu$ are determined self-consistently within the following two equations
\begin{eqnarray}
\label{eq10}\frac{\partial \la \mathcal{H} \ra}{\partial \mu} = 0, \\
\frac{\partial \la \mathcal{H} \ra}{\partial \pbar} = 0. 
\label{eq11}
\end{eqnarray}
The unit boson occupancy constraint is satisfied by Eq.~\ref{eq10} and the ground state energ
is minimized with respect to variational parameter $\pbar$ in Eq.~\ref{eq11}.

\section{Results \label{results}}

\begin{figure}
\centerline{\includegraphics[width=45mm]{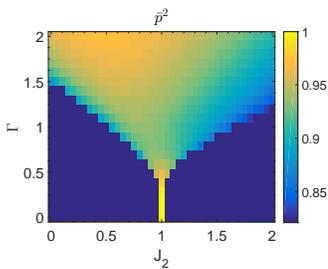}}
\caption{(color online) The density plot of $\pbar^2$, the probability of Bose-condensation of $u$=1-bosons on 
the plaquette background, as a function of $J_2$ and $\Gamma$. The plaquette 
order is strong in the bright region, while we do not get a consistent solution of Eqs.~\ref{eq10},~\ref{eq11} 
for dark blue regions, implying N\'{e}el and collinear states.}
\label{pdensity}
\end{figure}

First, we examine the validity region of POA, where the Bose-condensation of 
$u$=1-bosons has to appear. This is justified as far 
as $\pbar^2$ is close to unity, i.e. the strong plaquette order. 
The density plot of $\pbar^2$ versus $J_2$ and $\Gamma$ is shown in Fig.~\ref{pdensity}.
In the intermediate bright region of Fig.~\ref{pdensity}, we find $\pbar^2\gtrsim0.85$, 
which states that POA works very well. 
However, there exists dark blue area (in Fig.~\ref{pdensity}) in which we can not find a simultaneous solution 
of Eqs.~\ref{eq10},~\ref{eq11}. It shows that the unit boson occupancy constraint is not fulfilled in the 
dark blue regions. 
This seems to be a rational consequence of comparing the ground state energy of the 
primary plaquette background with those of the classical N\'{e}el or collinear backgrounds
(see Fig.~\ref{comparebackground}). Specially, for the low field and weakly frustrated regions
we get higher ground state energy for the plaquette background than the other classical backgrounds.
This is also a result of gap vanishing by approaching the dark blue area, where the 
elementary excitations of the model become gapless leading to a different type of ordering
with different background condensation.
Hence, approaching the dark blue region, the hypothesis of Bose-condensation 
of $u$=1-bosons can not be justified anymore.
However, we observe evidences for the existence 
of N\'{e}el and collinear phases by reaching the gapless critial border, which will be described in 
the following.

\begin{figure}
\centerline{\includegraphics[width=85mm]{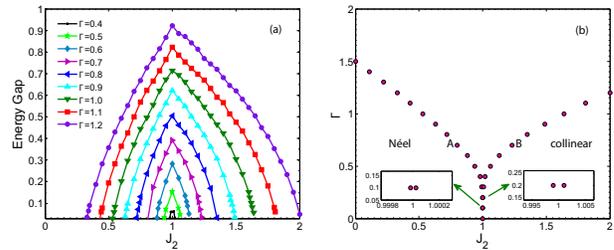}}
\caption{((color online) (a) The energy gap versus $J_2$ for different values of transverse field ($\Gamma$). 
The gap is finite in the intermediate region, while it vanishes at two critical values of $J_2$. 
(b) The location of critical points, which corresponds to the vanishing of energy gap in the phase diagram. 
The insets indicate an opening of a narrow region around $J_2=J_1$, where the gap is still finite for low fields.}
\label{borders}
\end{figure}

As a first indication, we evaluate the minimum of the excitation spectrum $\omega_{\bk,\nu}$, 
which defines the energy gap of our model. It is plotted in Fig.~\ref{borders}-(a) versus $J_2$ for
different values of the transverse field $\Gamma$. We observe that at a fixed value of $\Gamma$, the energy 
gap vanishes at two critical couplings of $J_2$, which corresponds to the locations of quantum phase transitions. 
%to either N\'{e}el or collinear states. 
The quantum critical points in the $\Gamma - J_2$ plane are shown in Fig.~\ref{borders}-(b), 
which displays the phase diagram of our model representing two critical boundaries.

\begin{figure}
\centerline{\includegraphics[width=75mm]{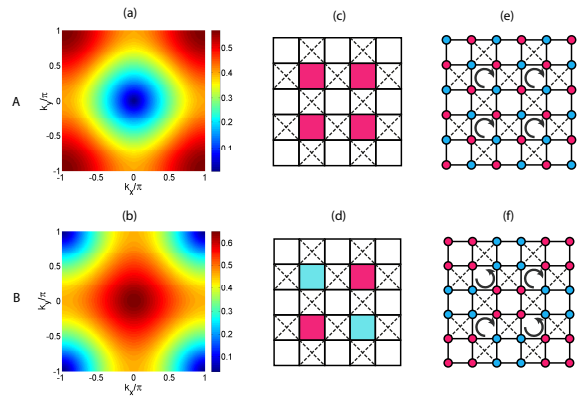}}
\caption{(color online) Up and Down rows correspond respectively to the critical points A ($J_2=0.80$) 
and B ($J_2=1.23$) for $\Gamma=0.7$. (a) and (b): Density plots of the lowest band of bosonic excitation spectrum. 
The excitation energy vanishes at 
($k_x=0, k_y=0$) for pint $A$ and ($\pm \pi, \pm \pi$) for point $B$. (c) and (d): the type of plaquette ordering of 
the groundstate. A symmetric covering for point $A$ and a staggered one for point $B$. (e) and (f): 
The classical representations for the N\'{e}el and collinear 
states, which can be mapped to the middle pictures (c) and (d) with plaquette-type orderings. Clockwise and 
counter-clockwise arrows represent the kind of arrangement of up and down (red and blue) spins in adjacent plaquettes.}
\label{excitationspectrum7} 
\end{figure}

One of the key features of the plaquette operator approach, compared to LSWT \cite{Henry:2012}, is to lift 
the exponential degeneracy of the classical collinear phase toward a unique quantum collinear state. 
Of course,
it leaves a fourfold degeneracy, two of them coming from the $Z_2$ symmetry and the 
other two from the translational symmetry. In order to demonstrate this assertion we 
have studied the lowest band of 
excitation spectrum $\omega_{\nu,\bk}$ for $\Gamma=0.7$ and at the two critical couplings $J_2=0.80$ 
and $J_2=1.23$, corresponding to the gapless points A and B, shown in Fig.~\ref{borders}-(b). 
The density plot of the lowest band of excitation spectrum is shown in Fig.~\ref{excitationspectrum7}-(a) and (b) corresponding
to the points A and B, respectively.

The density plots show that the gapless points occur at different k-vectors for A and B.
As we see, the excitation spectrum reaches a minimum at the ferromagnetic wave vector ($k_x=0, k_y=0$) 
for A (Fig.~\ref{excitationspectrum7}-(a)), 
while it becomes minimum at the anti-ferromagnetic wave vector ($k_x=\pm \pi, k_y=\pm \pi$) 
for B (Fig.~\ref{excitationspectrum7}-(b)). 
It reveals the construction of different orderings at these two critical points.
The wave vector ($k_x, k_y$) corresponds to the type of plaquette ordering of 
the lattice \cite{Note:1}. A minimum at the ferromagnetic wave vector ($k_x=0, k_y=0$) indicates a ferromagnetic tiling of 
resonating plaquettes, shown in Fig.~\ref{excitationspectrum7}-(c) 
which can be equivalent to a N\'{e}el configuration of the whole lattice shown 
in Fig.~\ref{excitationspectrum7}-(e). In fact, four shaded plaquettes in Fig.~\ref{excitationspectrum7}-(c) 
are in the same resonating state, similar to the same 
orientation of spins on neighboring plaquettes of a N\'{e}el state of Fig.~\ref{excitationspectrum7}-(e). 
Thus, the critical point A evidently 
expresses a transition to a N\'{e}el phase. 
On the other side, a minimum at ($k_x=\pm \pi, k_y=\pm \pi$) corresponds to a staggered (anti-ferromagnetic) 
plaquette covering of the lattice at point B which is equivalent to a specific collinear order for the whole 
lattice. In fact, the staggered ordering of plaquettes in Fig.~\ref{excitationspectrum7}-(d) acts 
like the opposite orientation of spins in the
adjacent  plaquettes of a particular collinear state shown in Fig.~\ref{excitationspectrum7}-(f). 
Hence, the critical point B corresponds to a quantum phase transition
to a unique collinear state. Interestingly, we conclude that plaquette-type excitations of POA are proper 
candidates to lift the extensive degeneracy of a classical groundstate, compared to the single-spin-flip ones of
LSWT \cite{Henry:2012}.

The above arguments are true for the whole transition points of Fig.\ref{borders}-(b),
i.e. the gap vanishes at ($k_x=0, k_y=0$) for the critical line in $J_2<1$ and
it vanishes at ($k_x=\pm \pi, k_y=\pm \pi$) for the critical line in $J_2>1$.
More justification for the N\'{e}el and collinear orders is given by the nearest neighbor (NN) and
next-nearest neighbor (NNN) correlation functions.
As shown in Fig.~\ref{correlations} (of \ref{ap-b}), the NN correlation $\langle S_i^z S_j^z \rangle_{\langle i,j\rangle}$ 
is always negative at both critical points A and B, while the NNN correlation $\langle S_i^z S_j^z\rangle_{\langle\langle i,j\rangle\rangle}$ is positive at 
point A and becomes negative for B, confirming the N\'{e}el and  collinear orders, respectively.
Accordingly, the two critical lines of the phase diagram correspond to a transition to the N\'{e}el 
phase for $J_2<1$ and to the collinear phase for $J_2>1$.

\begin{figure}
\centerline{\includegraphics[width=85mm]{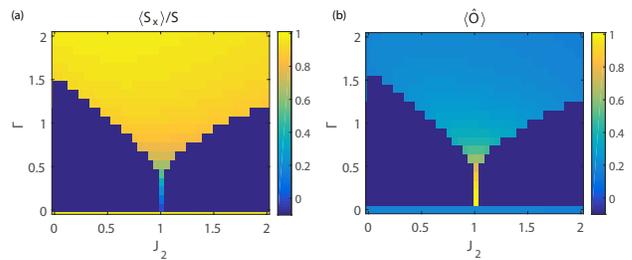}}
\caption{(color online) (a) The density plot of transverse magnetization $\langle S_x\rangle/S$ versus $J_2$ and $\Gamma$.
(b) The density plot of plaquette order parameter $\langle\hat{O}\rangle$ versus $J_2$ and $\Gamma$. The dark blue part is the region where the constraint of 
one boson per plaquette is not satisfied and contains no data.}
\label{orderparameters}
\end{figure}

Having determined the transition lines to the N\'{e}el and collinear phases, we now study the bright 
intermediate region of Fig.~\ref{pdensity}, in which we obtain the condensation of $u$=1-bosons ($\pbar^2\gtrsim0.85$). 
This area covers the regions, where LSWT breaks down in the vicinity of classical phase boundaries 
(see Fig.9 of Ref.~\cite{Henry:2012}). Thus, we expect that POA improves the phase diagram of the model, 
via plaquette-type quantum fluctuations. We calculate the transverse magnetization $\langle S_x\rangle$, 
which is shown as a density plot in Fig.~\ref{orderparameters}-(a) (for details see \ref{ap-c}).
The value of transverse magnetization is high enough
for $\Gamma\gtrsim0.3$, imitating a quantum paramagnet (polarized) phase. 
%It manifests that although 
%the plaquette operator approach is not able to reproduce the N\'{e}el and collinear long-range orders, 
%but it can reproduce the quantum paramagnet (polarized) phase. 
In fact, in the quantum paramagnet phase, $\langle S_x\rangle$ is less than its maximal classical value of $0.5$ due to 
strong quantum fluctuations. Thus, we conclude that POA can reproduce the polarized phase of high fields in such a way that all plaquettes are in their polarized states, restoring the translational symmetry of the lattice.

\begin{figure}[h]
\centerline{\includegraphics[width=95mm]{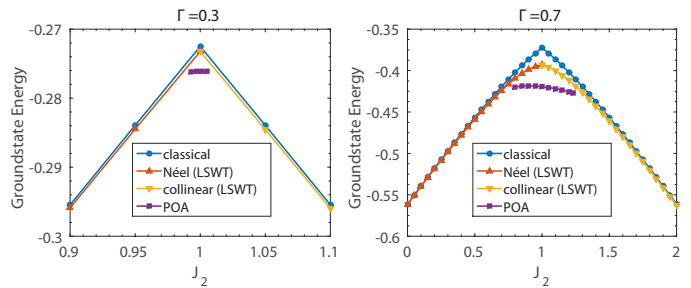}}
\caption{(color online) The groundstate energy per spin (in units of $J_1$) versus $J_2$ for $\Gamma=0.3$ and $0.7$. The upper blue line ($\bullet$) shows the result of classical ($S\rightarrow\infty$) approximation, 
the orange ({$\blacktriangle$}) and yellow ({$\blacktriangledown$}) lines represent the results of LSWT based on 
the N\'{e}el and collinear backgrounds, respectively (Ref.~\cite{Henry:2012}) and the lower purple 
line ({\tiny $\blacksquare$}) demonstrates the result of POA of the present work.}
\label{energy}
\end{figure}

On the other hand, for low transverse fields $\Gamma\lesssim0.3$, the value of transverse magnetization $\langle S_x\rangle$ 
deviates dramatically from its saturated value, revealing the onset of a new phase. 
In contrast to the result of LSWT \cite{Henry:2012},
the emergent new phase is 
neither a N\'{e}el nor a collinear state, which can be confirmed via 
energy considerations.
The ground state energy per spin (GSE) is plotted in Fig.~\ref{energy} versus $J_2$ 
for two different values of transverse field $\Gamma=0.3, 0.7$. 

Obviously, for both transverse fields and in the intermediate region around $J_2=J_1$, the GSE of POA is lower 
than the  corresponding classical and LSWT ones. It justifies strongly that POA gives a more precise representation of 
the groundstate for the bright region of Fig.~\ref{pdensity}. 
%We interpreted a quantum paramagnet phase 
%for $\Gamma\gtrsim0.4$, according to Fig.\ref{sx}, but what is the phase nature for low transverse 
%fields $\Gamma\lesssim0.4$? 
Now in order to understand the nature of phase for $\Gamma\lesssim0.3$, 
we define the resonating plaquette operator
\begin{equation}
 \hat{O}=\vert \varphi\rangle\langle {\bar \varphi}\vert+\vert {\bar \varphi}\rangle \langle \varphi \vert,
 \label{ohat}
\end{equation}
in which $\vert \varphi\rangle=\vert\uparrow\downarrow\uparrow\downarrow\rangle$ 
and $\vert {\bar \varphi} \rangle=\vert \downarrow\uparrow\downarrow\uparrow \rangle$ 
are two possible N\'{e}el configurations of a single plaquette ($\uparrow$ and $\downarrow$ 
represent the two eigenstates of $S^z$ operator at the four corners of a plaquette). 
In fact, $\hat{O}$ defines a measure of resonating magnitude between $\vert \varphi\rangle$ 
and $\vert {\bar \varphi} \rangle$ on a plaquette. 
% It is a suitable definition as it avoids 
%formation of magnetic long range orders like N\'{e}el and collinear states on the whole lattice. 
Hence, the expectation value of $\hat{O}$ is close to one for a resonating plaquette solid
state (RPS), which has no magnetic order in z-direction. Fig.~\ref{orderparameters}-(b) shows the density plot 
of $\langle\hat{O}\rangle$, which is an outcome of POA (for details see \ref{ap-c}). 
It is evident that for a narrow 
region around $J_2=J_1$ and $\Gamma\lesssim0.3$, the value of $\langle\hat{O}\rangle$ is very close to unity. 
However, there exist a small amount of field induced magnetization for this region (see Fig.~\ref{orderparameters}-(a))
that propose to call it a canted RPS phase.
It implies a resonating plaquette-type ordering in addition to a small 
inclination along the transverse field. A schematic representation of this phase is shown 
in Fig.~\ref{checkerboard}-(b). The emergent RPS phase breaks translational symmetry of 
the lattice and is two-fold degenerate. Therefore, the plaquette-type quantum fluctuations of 
POA are able to lift the extensive degeneracy of the square ice, leading to an {\it order by disorder}.

\begin{figure}
\centerline{\includegraphics[width=75mm]{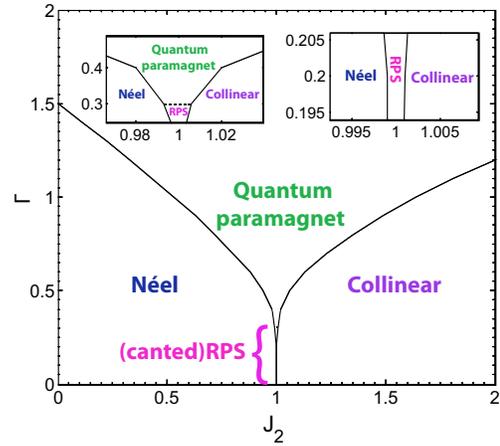}}
\caption{(color online) Phase diagram of S=1/2 TFIM on the checkerboard lattice within the plaquette operator approach. 
The phase boundaries to the N\'{e}el and collinear ordered states are denoted by black-solid lines.
Both insets show the narrow (canted) RPS phase, which fills the space between the N\'{e}el and
collinear phases around $J_2=J_1$ for $\Gamma\lesssim0.3$.}
\label{phasediagram}
\end{figure}

Finally, according to above arguments, the phase diagram of TFIM on the checkerboard lattice, 
obtained from POA, is sketched in Fig.~\ref{phasediagram}. In the limit of $J_2=0$, in which the system 
reduces to TFIM on the square lattice, the gap vanishes at $\Gamma=1.50$, which corresponds to the quantum 
phase transition from quantum paramagnet to the N\'{e}el phase.
It is in very good agreement with the results of density matrix renormalization group \cite{PhysRevB.57.8494}, extended coupled cluster method \cite{Rosenfeld:2000} and quantum Monte-Carlo simulation \cite{Blote:2002}, which report the critical point of the square lattice TFIM at $\Gamma=1.50$ \cite{Rosenfeld:2000} and $\Gamma=1.52$ \cite{Blote:2002,PhysRevB.57.8494}. This is a success of POA compared with LSWT \cite{Henry:2012}, which gives 
$\Gamma_c^{(LSWT)}=2.0$ for $J_2=0$. Thus, we anticipate that the whole critical lines
shown in Fig.~\ref{phasediagram} give an accurate phase diagram of the model. Moreover, the non-monotonic behavior of the ground state energy in LSWT leads to an inconsistency of
the sign of NNN correlation function close to the critical boundaries \cite{Henry:2012}.
In addition, we obtain an RPS state at low fields in a narrow region around $J_2=J_1$, 
which has not been observed via LSWT. The existence of a canted RPS phase confirms the results of Monte-Carlo studies of 
Refs.~\cite{Moessner:2004} and ~\cite{Henry:2014}, 
as well as the result of quantum dimer model \cite{Shannon:2004}.

As we mentioned earlier, the single boson occupancy constraint (Eq.~\ref{eq10}) is not satisfied 
in the area denoted by the N\'{e}el and collinear states in the phase diagram. Hence, we are not able to study the nature of phase transitions to these magnetically ordered phases, in terms of antiferromagnetic order parameter.
Nevertheless, we predict that transition from quantum paramagnet to either N\'{e}el or collinear 
phases should be of a continuous second order type, similar to the transition 
at $\Gamma=1.5$ for $J_2=0$ limit \cite{Rosenfeld:2000,Blote:2002,PhysRevB.57.8494}. 
%Accordingly, we also expect a continuous transition from the canted RPS to the N\'{e}el or collinear phases.

\begin{figure}
\centerline{\includegraphics[width=75mm]{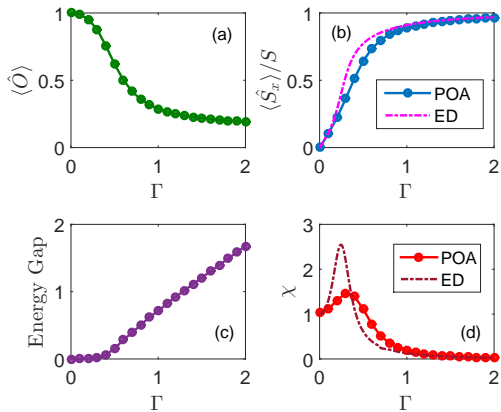}}
\caption{(color online) Features of the isotropic case $J_2=J_1$ versus $\Gamma$, (a) plaquette order 
parameter $\langle\hat{O}\rangle$, (b) transverse magnetization $\langle\hat{S_x}\rangle/S$ obtained from POA, which is compared with the result of a 16-sites Lanczos-ED calculation, (c) the energy gap which shows a linear behavior for $\Gamma\gtrsim0.3$, indicating a quantum paramagnet phase and (d) the first derivative of transverse magnetization with respect to $\Gamma$ ($\chi$), with a peak at $\Gamma\simeq0.3$ implying a phase transition. The data obtained from 16-sites Lanczos-ED calculation with a peak at $\Gamma\simeq0.24$  is also shown for comparison.}
\label{isotropic}
\end{figure}

Now, let us investigate the transition between RPS and quantum paramagnet phase at $J_2=J_1$. In this respect, 
we calculate some properties of the model at the isotropic case $J_2=J_1$ versus transverse field $\Gamma$, 
shown in Fig.~\ref{isotropic}. 
The plaquette order parameter $\langle\hat{O}\rangle$ is shown in Fig.~\ref{isotropic}-(a),
which indicates a deep decreasing from unity when increasing the transverse field $\Gamma$, 
although it does not reach zero because of strong quantum fluctuations. The reduction of plaquette order 
parameter is accompanied with a sharp increment of the transverse magnetization $\langle\hat{S_x}\rangle$ 
toward its saturated value, shown in Fig.~\ref{isotropic}-(b). 
Thus, there should be a continuous phase transition 
between RPS state at low fields and the quantum paramagnet phase for high fields. 
This is supported by the broken translational symmetry of the RPS phase compared with the
translational invariance of quantum paramagnet.
Nevertheless, the quasi-particle excitation gap, shown in Fig.~\ref{isotropic}-(c), vanishes only at $\Gamma=0$, which 
does not show a quantum phase transition at finite-$\Gamma$ and $J_2=J_1$.
However, the non-linear trend of energy gap for $\Gamma\lesssim0.3$ (see the inset of Fig.~\ref{isotropic}-(c)) is changed to the linear behavior for the $\Gamma\gtrsim0.3$, which is the property of a quantum paramagnet. 
On the other hand, the first derivative of transverse magnetization, i.e. the susceptibility, demonstrates a peak at $\Gamma\simeq0.3$ shown in Fig.~\ref{isotropic}-(d). To justify our results, we have employed the Lanczos exact-diagonalization method to calculate the ground state of our model. We consider a 16-sites ($4\times4$) lattice with periodic boundary condition. The results of magnetization ($\langle \hat{S}_x \rangle$) and its corresponding susceptibility have been shown in Fig.12 (b),(d) that show a fairly good agreement.

Accordingly, we deduce that a quantum phase transition should split the RPS and quantum paramagnet phases, although our approach does not show a zero gap mode, which should be checked precisely, using further numerical techniques. A similar situation has been observed in the POA of the frustrated honeycomb antiferromagnet, where the gap of POA does not vanish \cite{Ganesh:2013} at the expected transition points between a plaquette-RVB phase and N\'{e}el or dimer phases; However, numerical DMRG computations \cite{Ganesh:2013PRL}
justifies the closure of gap at the mentioned transition points.

\section{Summary and discussions \label{conclusion}}

We have studied the zero-temperature phase diagram of the transverse field Ising model on the checkerboard lattice, 
with nearest and next-nearest neighbor couplings $J_1$ and $J_2$, respectively. This model is a frustrated magnetic
system, which has an extensive degenerate classical groundstate for $J_2 \geq J_1$. The LSWT analysis of the model fails to lift 
the classical degeneracy of the collinear phase; moreover, the corresponding phase diagram 
show some instabilities near the classical 
boundaries \cite{Henry:2012}. This implies that the harmonic fluctuations, which come from the
single-spin-flip excitations of LSWT, can not give the true quantum fluctuations of the system, specially close to the
phase boundaries and highly frustrated region $J_2=J_1$.
Here, we have applied a plaquette operator 
approach, which is based on the bosonization of the model, in which a boson is associated to each plaquette eigenstate. A Bose-condensation of the plaquette ground
state is assumed, which survives as far as the excitation energy gap is non-zero.
The effective Hamiltonian, Eq.~\ref{eq9}, which takes into account the interaction between plaquettes,
describes the ground state phase diagram of the model. We would like to mention that the harmonic
fluctuations of the effective Hamiltonian are essentially an-harmonic fluctuations of the original 
spin model that are proper quantum fluctuations as the elementary excitations of the model.

\begin{figure}
\centerline{\includegraphics[width=50mm]{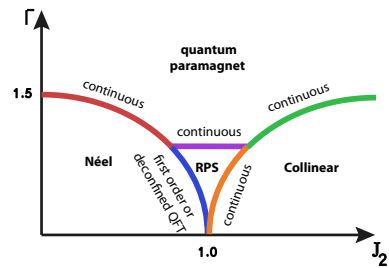}}
\caption{(color online) A schematic picture to specify the type of transitions on 
the phase diagram (Fig.~\ref{phasediagram}). The red and green lines correspond to 
continuous quantum phase transition (QPT) from quantum paramagnet to the N\'{e}el and collinear 
phases, respectively. The purple line shows the continuous transition from quantum 
paramagnet to RPS phase. The blue and orange lines are related to transitions from 
RPS phase to the N\'{e}el and collinear phases, respectively. The former should be 
either a first order type or a deconfined QPT, but the latter is a continuous QPT.}
\label{phasediagramcartoon}
\end{figure}

According to Fig.~\ref{comparebackground}, POA is a high-field approach, which also gives reliable results
for the low-fields in the highly frustrated region, where an emergent RPS phase shows up.
The phase diagram, Fig.~\ref{phasediagram}, consists of four phases,  quantum paramagnet phase, 
N\'{e}el phase, collinear ordered phase and an RPS phase for low fields $\Gamma\lesssim0.3$, 
a narrow region around $J_2=J_1$.
In fact, the exponential degeneracy of the 
classical groundstate at $J_2=J_1$ (square ice) is lifted toward a unique quantum RPS 
state that breaks translational 
symmetry of the lattice leaving two-fold degeneracy. 
It is a manifestaion of order-by-disorder that is induced by quantum fluctuations.
Disclosing the RPS phase is consistent with the results of the
quantum dimer model \cite{Shannon:2004} as an effective Hamiltonian on the degenerate Hilbert space
at $J_2=J_1$, which gives a plaquette ordered state for the zero chemical potential.
On the other hand, the boundaries to the N\'{e}el and collinear phases, corresponding to the
vanishing of quasi-particle excitation gap is determined consistently via POA, in contrast to LSWT. In addition, one of the smart features of POA is that the formation of N\'{e}el and collinear 
phases are realized according to the type of plaquette ordering at the transition points, which also reveals lifting the exponential degeneracy of the classical collinear phase toward a unique one.
Accordingly, the N\'{e}el and collinear phases are separated by a quantum paramagnet for the high-field region
and by an RPS phase for the low-field region, where the critical boundaries 
merge only at the zero field $\Gamma=0$.
Our POA results 
for $J_2=0$, manifest a transition from the N\'{e}el to quantum paramagnet 
at $\Gamma=1.50$, which is fully consistent with the result of TFIM on the square lattice with a 
second order phase transition at the critical field $\Gamma=1.50$ \cite{Rosenfeld:2000} 
or $\Gamma=1.52$ \cite{Blote:2002,PhysRevB.57.8494}. It suggests that the phase 
transition from quantum paramagnet to the N\'{e}el or collinear phases should be of a continuous 
second order type. This continuous phase transition persists for low fields between RPS and collinear phases. 
In fact, although both the collinear and RPS phases break translational symmetry, the $Z_2$ symmetry 
is only broken at the collinear phase, which suggests the transition to be of continuous type. 
However, the transition from RPS to the N\'{e}el phase at low fields should be a first order or 
a deconfined quantum phase transition \cite{Senthil:2004}, as they break different symmetries ($Z_2$ symmetry 
against translational symmetry).  Finally, we anticipate a continuous phase transition 
from RPS to quantum paramagnet phase (see Fig.~\ref{phasediagramcartoon}).

A recently Monte-Carlo study of the TFIM on the isotropic $J_2=J_1$ checkerboard lattice \cite{Henry:2014}, 
reports an RPS state, via an extrapolation to zero-temperature, that persists up to $\Gamma\simeq0.13$ and 
a canted N\'{e}el state for $0.13\lesssim\Gamma\lesssim0.28$ and finally a quantum paramagnet 
phase for higher fields ($\Gamma\gtrsim0.28$). However, the presence of such a N\'{e}el phase is 
a very delicate issue, which requires more justifications. According to Ref.~\cite{Henry:2014}, 
the N\'{e}el phase does not come from direct simulation on the origianl Hamiltonian, 
rather it is an outcome of the simulation on the fourth order effective Hamiltonian that 
can be constructed from the extensive degenerate manifold at $J_2=J_1$. 
Moreover, the extrapolated ($N\rightarrow \infty$) staggered magnetization is 
obtained to be $m_s \sim 10^{-2}$, which is very small. On the other hand, we do not observe 
a signature for a N\'{e}el phase at $J_2=J_1$, which convinces us that the phase diagram at the highly frustrated point $J_2=J_1$ consists only
of an RPS and a quantum paramagnet. Hence, we predict that a quantum phase transition should occur between the RPS and quantum 
paramagnet phases, although our approach does not show a zero gap mode, which 
can only be checked using precise numerical techniques.

%%%%%%%%%%%%%%%%%%%%%%%%%%%%%%%%%%%%%%%%%%%%%%%%%%%%%%%%%%%%%%%%%%%%%%%%%%%%%%%%%%%%%%%%%%%%%%
\section{Acknowledgment}
The authors would like to thank R. Moessner, I. Rousochatzakis and P. Thalmeier for fruitful discussions
and comments.
This work was supported in part by the Office of Vice-President for Research of
Sharif University of Technology.
A. L. gratefully acknowledges the Alexander von Humboldt Foundation for financial support.

\begin{appendix}
\renewcommand\thesection{Appendix \Alph{section}} 

\subsection{Ground-state energy for the RPS phase \label{ap-a}} 
\begin{figure}
\centerline{\includegraphics[width=35mm]{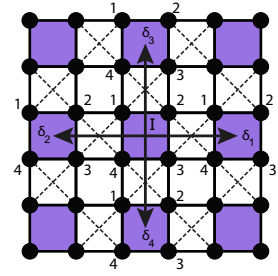}}
\caption{The checkerboard lattice: each isolated plaquette I interacts with four nearest-neighbor ones. 
The solid and dashed lines are $J_1$ and $J_2$ bonds, respectively.}
\label{appendixfig1}
\end{figure}

The interaction Hamiltonian between the isolated plaquette I and its four nearest-neighbor plaquettes reads 
as (see Fig.~\ref{appendixfig1}): 
\begin{eqnarray}
\nn H_{I\bm\delta_1}&=&J_1s_{2(I)}^zs_{1(I+\bm\delta_1)}^z+J_1s_{3(I)}^zs_{4(I+\bm\delta_1)}^z\\
\nn&&+J_2s_{2(I)}^zs_{4(I+\bm\delta_1)}^z+J_2s_{3(I)}^zs_{1(I+\bm\delta_1)}^z,          
\\ \nn H_{I\bm\delta_2}&=&J_1s_{1(I)}^zs_{2(I+\bm\delta_2)}^z+J_1s_{4(I)}^zs_{3(I+\bm\delta_2)}^z\\
\nn&&+J_2s_{4(I)}^zs_{2(I+\bm\delta_2)}^z+J_2s_{1(I)}^zs_{3(I+\bm\delta_2)}^z,\\
\nn H_{I\bm\delta_3}&=&J_1s_{1(I)}^zs_{4(I+\bm\delta_3)}^z+J_1s_{2(I)}^zs_{3(I+\bm\delta_3)}^z\\
\nn&&+J_2s_{1(I)}^zs_{3(I+\bm\delta_3)}^z+J_2s_{2(I)}^zs_{4(I+\bm\delta_3)}^z,\\ 
\nn H_{I\bm\delta_4}&=&J_1s_{4(I)}^zs_{1(I+\bm\delta_4)}^z+J_1s_{3(I)}^zs_{2(I+\bm\delta_4)}^z\\
\nn&&+J_2s_{3(I)}^zs_{1(I+\bm\delta_4)}^z+J_2s_{4(I)}^zs_{2(I+\bm\delta_4)}^z.\\
\end{eqnarray}
Accordingly, we obtain an effective Hamiltonian for the plaquette-ordered background of POA :
\begin{eqnarray}
\nn\mathcal{H}&=&N\pbar^2(\epsilon_1-\mu)+ N\mu+\sum_{I}\sum_{u}(\epsilon_{u}-\mu)b_{I,u}^\dg b_{I,u}\\
\nn &&+\frac{1}{2}\pbar^2\sum_{I}\sum_{\bm\delta=\bm\delta_1}^{\bm\delta_4}\sum_{u,v}[\la uv\vert H_{I\bm\delta} \vert11\ra b_{I,u}^\dg b_{I+\bm\delta,v}^\dg\\
\nn&&+\la u1\vert H_{I\bm\delta}\vert1v\ra  b_{I,u}^\dg b_{I+\bm\delta,v}+H.C.],
\\
\label{eqa2}
\end{eqnarray}
where index $I$ runs over all shaded plaquettes and $\bm\delta$ sums 
over the four nearest neighbors of each plaquette as shown in Fig.~\ref{appendixfig1}.

\begin{table*}
\caption{The values of transition amplitude $\la u\vert s_{\alpha}^z \vert1 \ra$, from the groundstate of a single 
plaquette to its sixteen eigenstates, for different values of transverse field $\Gamma$.}
\label{tab}       % Give a unique label
% For LaTeX tables use
\begin{tabular}{ccccccccccccccccc}
\hline\noalign{\smallskip}
u&1&2&3&4&5&6&7&8&9&10&11&12&13&14&15&16  \\
\noalign{\smallskip}\hline\noalign{\smallskip}
&&&&&&&&$\Gamma=1.0$&\\
\noalign{\smallskip}\hline\noalign{\smallskip}
$\la u\vert s_{1}^z \vert1 \ra$&0& -0.332& 0& -0.318& -0.194& 0& 0& 0& 0& 0& 0.012& 0& 0& -0.010& -0.002& 0\\
$\la u\vert s_{2}^z \vert1 \ra$&0& 0.332& -0.318& 0& -0.194& 0& 0& 0& 0& 0& -0.012& 0& -0.010& 0& -0.002& 0\\
$\la u\vert s_{3}^z \vert1 \ra$&0& -0.332& 0& 0.318& -0.194& 0& 0& 0& 0& 0& 0.012& 0& 0& 0.010& -0.002& 0\\
$\la u\vert s_{4}^z \vert1 \ra$&0& 0.332& 0.318& 0& -0.194& 0& 0& 0& 0& 0& -0.012& 0& 0.010& 0& -0.002& 0\\
\noalign{\smallskip}\hline\noalign{\smallskip}
&&&&&&&&$\Gamma=0.5$&\\
\noalign{\smallskip}\hline\noalign{\smallskip}
$\la u\vert s_{1}^z \vert1 \ra$&0& -0.424& 0& -0.220& -0.136& 0& 0& 0& 0& 0& 0.040& 0& 0& -0.029& -0.002& 0\\
$\la u\vert s_{2}^z \vert1 \ra$&0& 0.424& -0.220& 0& -0.136& 0& 0& 0& 0& 0& -0.040& 0& -0.029& 0& -0.002& 0\\
$\la u\vert s_{3}^z \vert1 \ra$&0& -0.424& 0& 0.220& -0.136& 0& 0& 0& 0& 0& 0.040& 0& 0& 0.029& -0.002& 0\\
$\la u\vert s_{4}^z \vert1 \ra$&0& 0.424& 0.220& 0& -0.136& 0& 0& 0& 0& 0& -0.040& 0& 0.029& 0& -0.002& 0\\
\noalign{\smallskip}\hline\noalign{\smallskip}
&&&&&&&&$\Gamma=0.1$&\\
\noalign{\smallskip}\hline\noalign{\smallskip}
$\la u\vert s_{1}^z \vert1 \ra$&0& -0.497& 0& -0.030& -0.025& 0& 0& 0& 0& 0& 0.023& 0& 0& -0.020& -0.000& 0\\
$\la u\vert s_{2}^z \vert1 \ra$&0& 0.497& -0.030& 0& -0.025& 0& 0& 0& 0& 0& -0.023& 0& -0.020& 0& -0.000& 0\\
$\la u\vert s_{3}^z \vert1 \ra$&0& -0.497& 0& 0.030& -0.025& 0& 0& 0& 0& 0& 0.023& 0& 0& 0.020& -0.000& 0\\
$\la u\vert s_{4}^z \vert1 \ra$&0& 0.497& 0.030& 0& -0.025& 0& 0& 0& 0& 0& -0.023& 0& 0.020& 0& -0.000& 0\\
\noalign{\smallskip}\hline
\end{tabular}
\end{table*}
A transition-amplitude like $\la uv\vert J_1s_{2(I)}^zs_{1(I+\bm\delta_1)}^z \vert11 \ra$ 
can be reduced to a product of matrix elements of the
single-plaquette operators as:
\begin {eqnarray}
\la uv\vert J_1s_{2(I)}^zs_{1(I+\bm\delta_1)}^z \vert11 \ra=J_1 \la u\vert s_{2(I)}^z \vert1 \ra \times 
\nn\la v\vert s_{1(I+\bm\delta_1)}^z \vert1 \ra.\\
\label{A1}
\end{eqnarray}
We have plotted the transition matrix elements versus magnetic field in Fig.~\ref{transition}, however, we summarize
few cases in the following table to have an impression of the values.
Table.~\ref{tab} shows the matrix elements $\la u\vert s_{\alpha}^z \vert1 \ra$ 
for three values of transverse field $\Gamma=1.0,0.5,0.1$ in which $u$ runs over 
sixteen eigenstates of a single plaquette and $\alpha=1,2,3,4$ represents the four spin-z operators at the four 
corners of a plaquette. 
It reveals that transition from 
groundstate $\vert 1 \rangle$ to eight eigenstates $\vert u \rangle$ of a single plaquette is non-zero, 
within which only four of them have a significant value, corresponding to $u=2,3,4,5$. 
However, we observe that for low values of transverse field $\Gamma$, the transition to the 
first excited state $u=2$ is more dominant. Therefore, for low transverse fields in which we obtain an emergent 
RPS phase, we can reduce the number of excited bosons that contribute to the effective Hamiltonian, 
to only 'one' boson i.e. the first excited $u=2$ state of each plaquette. Accordingly, in the following we only 
consider the $u=2$ state as an excited boson, participating in the effective Hamiltonian of Eq.~\ref{eqa2}. 
We calculate analytically the groundstate energy of RPS phase, as well as spin-spin correlation functions and order parameters via 
this simplified version of POA.

In order to diagonalize the effective Hamiltonian, we first rewrite it in the momentum space representation
using the following transformations
\begin{eqnarray}
b_{\bk,u} = \frac{1}{\sqrt{N}}\sum_{\bk} b_{I,u}e^{-i\bk.\bm r_I},\hspace{1em} H_{\bk}=\sum_{\bk} H_{\bm\delta}e^{i\bk.\bm\delta},
\end{eqnarray}
which gives
\begin{eqnarray}
\mathcal{H} &=& N\pbar^2(\epsilon_1- \mu) + N\mu-\frac{1}{2}N (\epsilon_{2}-\mu)  \nn \\
&&+ \frac{1}{2}\sum_{\kb}(b_{\kb,2}^\dg,b_{-\kb,2})\bm M_{\kb}(b_{\kb,2},b_{-\kb,2}^\dg)^T ,
\label{eqp}
\end{eqnarray}
where, the groundstate energy of a plaquette ($\epsilon_1$), and the first excited one ($\epsilon_2$) 
have the following expressions
\begin{eqnarray}
\epsilon_1=-\frac{\sqrt{1+4\Gamma^2+\sqrt{1+16\Gamma^4}}}{\sqrt{2}},
\nn\epsilon_2=\frac{1}{2}(-1-\sqrt{1+4\Gamma^2}).\\
\label{eq6}
\end{eqnarray}
The elements of $\bm M_{\bk}$ is as follows,
\begin{eqnarray}
\nn M_{11}&=&M_{22}=(\epsilon_2-\mu)+2f\pbar^2(cosk_x+cosk_y),\\
\nn M_{12}&=&M_{21}=2f\pbar^2(cosk_x+cosk_y),\\
\end{eqnarray}
in which $f$ is a function of $\Gamma$ and $J_2$ (which has not been shown here due to its
long expression). The Hamiltonian Eq.~\ref{eqp} is diagonalized via a paraunitary Bogoliubov transformation \cite{Colpa:1978} as,
\begin{eqnarray}
\mathcal{H}=N\pbar^2(\epsilon_1- \mu) + N\mu-\frac{1}{2} N(\epsilon_2-\mu)
\nn +\sum_{\bk}(\frac{1}{2}+\gamma^\dagger_{\bk}\gamma_{\bk})\omega_{\bk},\\
\end{eqnarray}
where the eigenmodes read as 
\begin{eqnarray}
\nn\omega_{\bk}=\sqrt{(\epsilon_2-\mu)(\epsilon_2-\mu+4f\pbar^2(cosk_x+cosk_y)}.\\
\end{eqnarray}
Finally, the groundstate energy of the RPS phase becomes
\begin{eqnarray}
\nn E_{RPS}=N\pbar^2(\epsilon_1- \mu) + N\mu-\frac{1}{2} N(\epsilon_2-\mu)+\frac{1}{2}\sum_{\bk}\omega_{\bk},\\
\label{eqA10}
\end{eqnarray}
in which $\bar{p}$ and $\mu$ are determined self-consistently using simultaneous numerical solution of the following equations
\begin{eqnarray}
\frac{\partial E_ {RPS}}{\partial\mu} = 0, \label{eqA11}\\
\frac{\partial E_{RPS} }{\partial\pbar} = 0. \label{eqA12}
\end{eqnarray}
Eq.~\ref{eqA11} satisfies the unit boson occupancy constraint, and Eq.~\ref{eqA12} minimizes the 
ground state energy with respect to $\pbar$.

\subsection{Nearest and next-nearest neighbor correlation functions \label{ap-b}}
\begin{figure}
\centerline{\includegraphics[width=55mm]{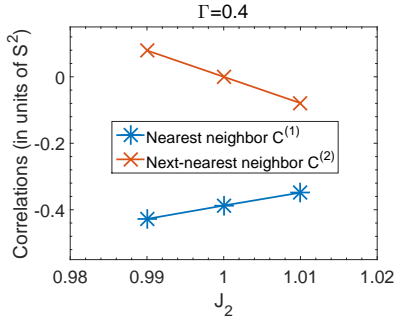}}
\caption{(color online) The nearest and next-nearest neighbor correlations as a function of $J_2$, for a low transverse field $\Gamma=0.4$. The change of sign at $J_2=1$ for the next-nearest neighbor correlation function represents the change of tendency to construct different orderings, i.e. a N\'{e}el order for $J_2\lesssim0.99$ 
and a collinear order for $J_2\gtrsim1.01$.}
\label{correlations}
\end{figure}

The nearest and next-nearest neighbor correlation functions corresponding to 
$C^{(1)}=\langle S_{i}^{z}S_{j}^{z}\rangle_{\langle i,j\rangle}=\partial{\langle \mathcal{H}\rangle}/\partial{J_{1}}$ 
and $C^{(2)}=\langle S_{i}^{z}S_{j}^{z}\rangle_{\langle\langle i,j\rangle\rangle}=\partial{\langle \mathcal{H}\rangle}/\partial{J_{2}}$ are plotted in Fig.~\ref{correlations} versus $J_2$, at a low transverse field $\Gamma=0.4$. The plot justifies the change of ordering structure to the N\'{e}el and collinear phases, when approaching to the critical points $J_2\approx0.99$ and $J_2\approx1.01$, respectively. As a matter of fact, it is expected from the classical picture presented in Fig.~\ref{checkerboard} that the nearest neighbor correlations in a collinear phase is weaker, in strength, than in the N\'eel phase (the average number of aligned and anti-aligned 
nearest-neighbor bonds are roughly the same in the collinear order, compared to all anti-aligned nearest-neighbor 
bonds in the N\'{e}el order). This is confirmed evidently in the plot of nearest neighbor correlation of 
Fig.~\ref{correlations}. Moreover, according to the classical picture, the next-nearest neighbor correlations are 
positive for the N\'{e}el order, while they are negative in the collinear order. Therefore, the change of sign in the next-nearest-neighbor correlation at $J_2=1$ in Fig.~\ref{correlations} is a signature of entering from the RPS phase to the N\'{e}el and collinear phases, at $J_2\approx0.99$ and $J_2\approx1.01$, respectively.

\subsection{Order Parameters \label{ap-c}}

The expectation values of the order-parameter operators $\hat{O}$ and $\hat{S_x}$ are calculated for 
the RPS state at low fields, making use of the density matrix formalism. In the simplified one-excited-boson 
version of POA, the density matrix operator $\hat{\rho}$ of a single plaquette takes the following form
\begin{eqnarray}
\hat{\rho}=\pbar^2\vert{1}\rangle\langle{1}\vert+(1-\pbar^2)\vert{2}\rangle\langle{2}\vert
\end{eqnarray}
where $\vert{1}\rangle$ denotes the groundstate of the single plaquette, $\vert{2}\rangle$ is the first 
excited state, and $\pbar^2$ is the probability of finding the 
single plaquette in its groundstate.

The expectation values of the plaquette order parameter $\hat{O}$ (which is defined in Eq.~\ref{ohat}) and the transverse magnetization are given by the following equations
\begin{eqnarray}
\label{B5}\nn\langle\hat{O}\rangle&=&Tr(\hat{O}\hat{\rho})=\pbar^2\langle{1}\vert \hat{O}\vert{1}\rangle+
(1-\pbar^2)\langle{2}\vert\hat{O}\vert{2}\rangle,\\
\nn\langle\hat{S_x}\rangle&=&Tr(\hat{S_x}\hat{\rho})=\pbar^2\langle{1}\vert \hat{S_x}\vert{1}\rangle+
(1-\pbar^2)\langle{2}\vert\hat{S_x}\vert{2}\rangle.\\
\end{eqnarray}

Extending the above arguments to the general version of POA (with four excited bosons, contributing to the effective Hamiltonian), 
we obtain the groundstate energy, the order parameters and other quantities for the whole values of transverse field $\Gamma$, as presented in Sec.~\ref{results}.
\end{appendix}

\bibliography{ITF-checkerboard}

\begin{thebibliography}{46}
\bibitem [{Not()}]{Note:1}%
  \BibitemOpen
  \href@noop {} {}\bibinfo {note} {We remind that according to the plaquette
  background of POA, the $k$-space vector runs over the first brillouin zone of
  an square lattice (set its lattice spacing to 1), constructed from the
  centers of isolated plaquettes (one every two uncrossed squares) of the
  checkerboard lattice.}\BibitemShut {Stop}%
\end{thebibliography}
\end{document}